\documentclass[11pt]{article}

\usepackage[final]{acl}

\usepackage{times}
\usepackage{latexsym}
\usepackage{enumitem}
\usepackage[T1]{fontenc}
\usepackage{pifont}
\usepackage[utf8]{inputenc}

\usepackage{microtype}

\usepackage{inconsolata}

\usepackage{graphicx}
\usepackage{multirow}
\usepackage{booktabs}
\usepackage{amsmath}
\usepackage{amssymb}
\usepackage{comment}
\newcommand{\best}[1]{\textbf{#1}}
\newcommand{\gain}[1]{\textcolor{ForestGreen}{\textbf{#1}}}
\newcommand{\weakgain}[1]{\textcolor{gray}{#1}}

\usepackage{listings}
\definecolor{ForestGreen}{RGB}{34, 139, 34}

\definecolor{backcolour}{RGB}{245,245,245}
\definecolor{codegray}{rgb}{0.5,0.5,0.5}
\definecolor{codepurple}{rgb}{0.58,0,0.82}
\usepackage{pifont}

\lstdefinestyle{promptstyle}{
    backgroundcolor=\color{backcolour},
    basicstyle=\ttfamily\footnotesize,
    breaklines=true,
    keepspaces=true,
    frame=single,
    showspaces=false,
    showstringspaces=false,
    showtabs=false,
    tabsize=2
}

\title{Can Coding Agents Solve Repository-Level Issues with Rendered Code? An Exploratory Study of Visual Representations}

\author{
Weijie Liang\textsuperscript{1},
Yuanfeng Song\textsuperscript{2},
Xing Chen\textsuperscript{2},
Caleb Chen Cao\textsuperscript{1},
Sirui Han\textsuperscript{1},
Yike Guo\textsuperscript{1}
\\
\textsuperscript{1}The Hong Kong University of Science and Technology, Hong Kong, China\\
\textsuperscript{2}ByteDance, China
}

\begin{document}
\maketitle
\begin{abstract}
Visual modality has recently been explored as a way to compress textual tokens,
including rendering code as images for static code understanding. We study
whether this representation can serve as operational context for
\textit{agentic coding}, where an agent must navigate repositories, edit source
files, and verify executable patches. Using SWE-bench Verified, we evaluate rendered code in repository-level repair workflows and introduce controlled agent settings to separate unguided repository exploration from more structured repair stages.
Our results show a mixed picture. Rendered code consistently reduces
prompt-token cost, but the savings do not increase linearly with the nominal
visual compression ratio. It largely preserves end-to-end repair accuracy, but
does not overcome the performance limits of the underlying model or agent
architecture, and can become unstable under aggressive compression. Further
analysis suggests that visual code is most useful when raw source reading is a
major bottleneck; once repository localization is structured, much of the
remaining cost comes from patch--test trial-and-error, where visual compression
has limited leverage. Overall, our study positions rendered code as a viable
but conditional compression mechanism for realistic coding agents.
\end{abstract}

\begin{figure*}
    \centering
    \includegraphics[width=0.95\linewidth]{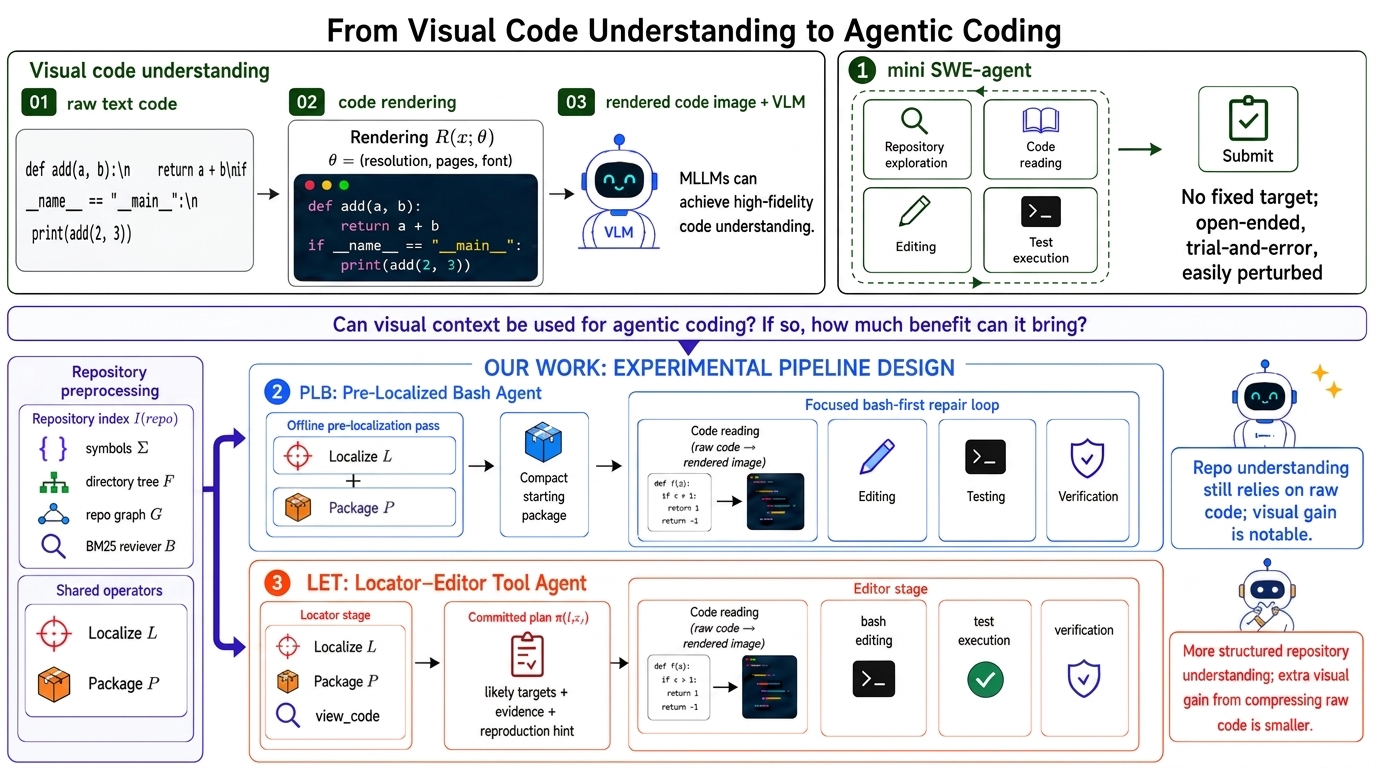}
    \caption{\textbf{Pipeline comparison between passive visual code understanding and agentic coding workflows.} The top panels contrast existing code rendering/VLM understanding with Panel \ding{182} the unguided mini SWE-agent task workflow. The bottom panels illustrate our experimental pipeline featuring two distinct agent paradigms: \ding{183} PLB (Pre-Localized Bash Agent), which uses an offline localization pass before a focused repair loop, and \ding{184} LET (Locator-Editor Tool Agent), which dynamically splits repository actions into explicit Locator and Editor stages to systematically analyze the utility and limitations of rendered code context.}
    \label{fig:intro}
    \vspace{-15pt}
\end{figure*}
\section{Introduction}

In recent years, with the development of long-context models and agentic systems~\citep{yao2023react,guo2024large}, how to efficiently represent and utilize large amounts of textual information has become an increasingly important problem. Traditional approaches usually rely on retrieval, summarization, or selective compression, while recent line of work has begun to explore converting textual information into
visual forms, treating images as a compact carrier for textual context
\citep{deepseekocr,Glyph,VTC-R1,Render-of-Thought,MemOCR,liang2026vizomem}.

In the coding domain, this idea has also been studied by rendering source code
into visual representations for code understanding and long-context code
compression \citep{codeocr,longcodeocr}.
However, existing studies mostly evaluate rendered code in static understanding
settings, where the model is given code to ``read''.
Agentic coding is different: \textit{real repair agents must search across files, localize relevant regions, edit code, run tests, and revise hypotheses over long interactive trajectories.}
This leaves a practical gap: success in rendered-code understanding does not by
itself show that visual representations can serve as effective working context
for coding agents.

To bridge the above-mentioned gap, we pioneer a comprehensive exploratory study
of rendered code as an alternative representation to raw code in real agentic
coding tasks. More precisely, we evaluate visual code inside repository-level
repair workflows, where agents must search the codebase, inspect relevant
context, edit source files, and verify executable patches through test
execution. 

Our exploratory study makes the following contributions and findings:
\begin{itemize}[nosep, leftmargin=*]
    \item \textbf{We provide the first systematic study of rendered code as operational context for agentic coding.}
    Prior work has mainly evaluated rendered code as static input for passive code understanding. We move this question into repository-level repair and find that
rendered code can substantially reduce token cost, but its realized savings are
constrained by code structure and readability, and therefore do not scale
linearly with the nominal compression ratio. This contrasts with the smoother
scaling behavior often assumed or empirically observed in general visual-text
compression settings~\citep{deepseekocr}.

    \item \textbf{We introduce a controlled evaluation framework to study how visual modality affects autonomous repository repair agents.}
As overviewed in Figure~\ref{fig:intro}, beyond directly testing rendered code
in mini SWE-agent workflows \cite{yang2024sweagent}, we design two controlled agent variants, PLB
(Pre-Localized Bash Agent) and LET (Locator-Editor Tool Agent).
These configurations progressively decouple repository exploration,
localization, source inspection, and editing, allowing us to trace where the
effects of rendered code are actually realized.

    \item \textbf{We identify three mechanisms that shape agentic repair with rendered code.}
    First, initial context reshapes repair behavior by compressing unguided
    exploration into the early trajectory and shifting later actions toward
    focused test--edit repair. Second, visual compression is most useful when raw
    source reading remains a major cost, and its marginal benefit shrinks once
    structured repository information has already reduced source-reading demand.
    Third, downstream trial-and-error matters more than repository understanding
    alone: even with better localization, final resolution is often dominated by
    editing, testing, and patch refinement.
\end{itemize}

\section{Related Work}


\noindent \textbf{Context Compression and Visual Information Representation.}
The computational cost of Transformer-style language models grows with the
number of input tokens, making context representation a central bottleneck for
long-document and interactive tasks \citep{vaswani2017attention}. A large body of work therefore studies how to expose only the most useful information to the model, with retrieval-augmented generation and related retrieval-centric methods being the most representative direction \citep{lewis2020retrieval, karpukhin2020dense, izacard2021leveraging, guu2020realm}. Another line compresses textual context through prompt-level rewriting, including summarization, learned compression, and selective token retention \citep{xu2024recomp, jiang2023llmlingua, jiang2024longllmlingua, li2023selective}. In this sense, they shorten context by transforming the original textual information.

Recent work explores a complementary route: changing the representation modality itself, using visual inputs as a dense carrier for the same textual information. DeepSeek-OCR
\citep{deepseekocr} studies optical 2D mappings for compressing text into
images, while Glyph \citep{Glyph} shows that rendered text can replace part of
raw textual context in long-text question answering. 

Other systems push this
idea beyond document compression. VTC-R1 \citep{VTC-R1} incorporates visual
text compression into a reasoning pipeline, while Render-of-Thought
\citep{Render-of-Thought} renders intermediate reasoning traces so that
chain-of-thought can be stored and consumed through the visual channel.
MemOCR \citep{MemOCR} and VizoMem \citep{liang2026vizomem} explore visual memory representations for organizing accumulated long-horizon information.
This perspective naturally extends to the coding domain, where source code constitutes a structured textual-symbolic artifact that can be transformed into visual representations.
\smallskip

\noindent \textbf{Coding Agents for Repository-Level Software Engineering.}
Repository-level coding benchmarks have shifted code generation from isolated
function synthesis toward realistic software maintenance. SWE-bench and
SWE-bench Verified evaluate agents on real GitHub issues, requiring them to
localize repository context, edit the implementation, and pass
project tests \citep{jimenez2024swebench, chowdhury2024swebenchverified}.
This setting has motivated coding agents and repair systems that combine
language models with repository search, shell interaction, editing, and test
feedback, including SWE-agent, AutoCodeRover, Agentless, and broader software
engineering agent frameworks such as OpenHands
\citep{yang2024sweagent, zhang2024autocoderover, xia2024agentless,
wang2025openhands}. In parallel, CodeOCR \citep{codeocr} and LongCodeOCR \citep{longcodeocr} explore rendered code as a visual representation of source code: by controlling image resolution, code can be directly compressed in the visual space while vision-language models still retain code understanding ability. 

However, understanding rendered code remains a largely passive setting, while agentic repair requires the model to actively acquire context by deciding which files and regions to inspect, use the resulting evidence to edit code and test over long interactive trajectories. This leaves a practical gap between rendered-code understanding and agentic software repair. To our knowledge, our work is the first to study whether visually rendered code can serve as operational context for repository-level coding agents, and where its effect appears across the repair procedure.

\begin{figure}[t]
    \centering
    \includegraphics[width=\linewidth]{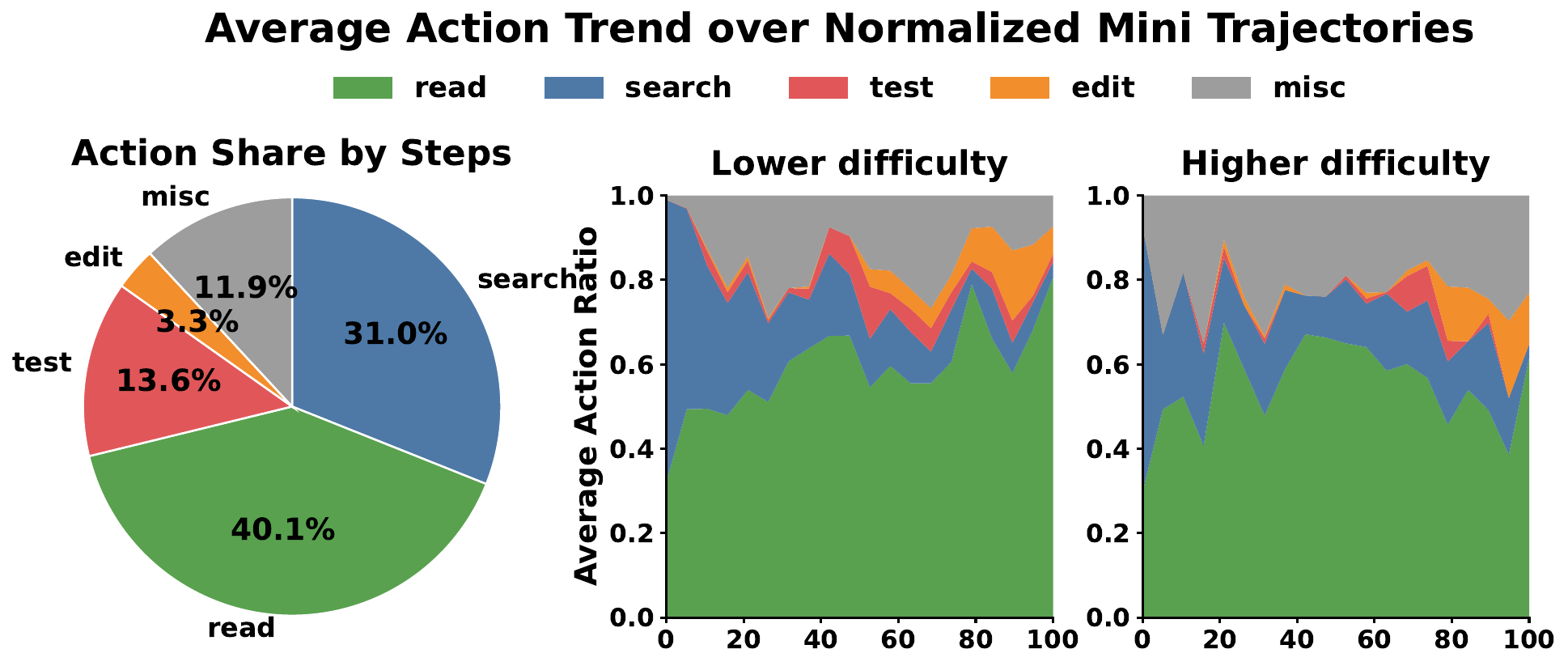}
    \caption{
   Action distribution and temporal action trends of mini SWE-agent v2 on SWE-bench Verified. Read/search actions dominate across both lower- and higher-difficulty tasks, and remain distributed throughout the normalized repair process. Difficulty groups are defined by the steps used to solve each task.
    }
    \label{fig:mini-agent-actions}
    \vspace{-15pt}
\end{figure}

\section{From Visual Code Understanding to Repository-Level Repair}
\label{sec:visual-agentic-bridge}

CodeOCR and LongCodeOCR established that VLMs can understand rendered code in isolation. Agentic coding is a different regime: rendered code must serve as a working context inside an interactive, repository-level workflow, where the agent navigates, edits, and verifies patches via test execution. The question we study is whether visual code retains its benefits once it is embedded in this loop, and where those benefits are actually realized.

We approach this in three layers. First, we set up the agentic task and identify the attribution challenge that evaluation must face (\S\ref{subsec:setup}). Second, we examine the rendering pipeline itself and show that its cost surface is fundamentally non-smooth in the nominal compression ratio (\S\ref{subsec:cost-surface}). With both pieces in hand, we then introduce the shared protocol and controlled progression of the agent configurations used in the main study (\S\ref{sec:experimental-setup}).

\subsection{Setup: Agentic Coding and the Attribution Challenge}
\label{subsec:setup}

A repository-level repair instance is defined by an issue description $I$ and a repository $\mathcal{R}$ consisting of source files, tests, and project metadata.
Given $(I,\mathcal{R})$, the agent must inspect the repository and produce a patch
$P = \pi(I, \mathcal{R})$
that passes both fail-to-pass tests $\mathcal{T}_{\mathrm{F2P}}$ and pass-to-pass tests $\mathcal{T}_{\mathrm{P2P}}$:
\begin{equation}
\label{eq:swebench-eval}
\mathrm{Pass}(\mathcal{R} \oplus P, \mathcal{T}_{\mathrm{F2P}})
\land
\mathrm{Pass}(\mathcal{R} \oplus P, \mathcal{T}_{\mathrm{P2P}}).
\end{equation}

The default solving paradigm is the bash-based mini SWE-agent (we adopt the v2 version), a lightweight agent that interacts with repositories via shell commands and mixes repository search, file inspection, code editing, test execution, and feedback inspection inside a single command-line trajectory. Figure~\ref{fig:mini-agent-actions} shows how the five action types are distributed across the task trajectory; details are provided in Appendix~\ref{appendix:action-types}. Two observations matter for our purposes. First, read and search actions dominate the trajectory. Second, code reading is distributed across the entire normalized repair process rather than confined to an initial localization stage, i.e., the agent repeatedly performs exploratory \texttt{search + read} interleaved with editing, testing, and feedback inspection.

This creates an attribution challenge for visual evaluation. Visual compression changes only how raw code is represented, while a bash trajectory also contains substantial non-code context, such as search outputs, test results, and patch feedback. Replacing textual code views with rendered images therefore perturbs only part of the interaction history, and the exploration loop may amplify or dilute this effect. A single end-to-end text-to-visual swap cannot tell whether observed changes come from the visual modality itself, trajectory variance, or their interaction. This is the gap between visual code understanding and repository-level agentic tasks: the model is no longer evaluated on a clean reading event, and the modality effect becomes entangled with search, editing, and verification.

\subsection{Analysis of the Visual Code Cost Surface}
    \label{subsec:cost-surface}

We first describe the rendering pipeline, since its cost--output behavior
  determines what any visual experiment can ultimately reveal.

  Let $x$ denote a code window and $r \in \mathbb{R}_{>0}$ a nominal compression
  ratio. The renderer maps $(x,r)$ to a target visual budget and searches over a
  finite set of layout configurations $\theta \in \Theta$, where each
  configuration specifies properties such as resolution, pagination, and font
  size. The selected layout
  \[
  \theta^{*}(x,r)
  =
  \arg\min_{\theta\in\Theta}
  \mathcal{L}\!\bigl(R(x;\theta),\,B(x,r)\bigr)
  \]
  balances target-budget matching against layout efficiency and page overhead.

  Here, $B(x,r)$ denotes the renderer-side target budget induced by code window
  $x$ and nominal ratio $r$, while $R(x;\theta)$ denotes the rendered visual
  output under layout configuration $\theta$. Detailed rendering settings, search space, and implementation choices are given in Appendix~\ref{app:rendering}. However, this code rendering pipeline has several intrinsic properties that directly affect its cost behavior.
  
  \noindent\textbf{(Property 1) Discrete realizability:}
  Although the nominal budget varies continuously, the renderer searches over a
  discrete candidate set $\Theta$. As a result, many nearby budgets collapse to
  the same rendered layout and hence the same realized cost.

  \noindent\textbf{(Property 2) Piecewise-constant tokenization:}
  Image-token cost changes only when a rendered image crosses a patch or tile
  boundary. Once a layout falls into the minimum active token bucket, further
  changes in resolution may degrade readability without changing token cost.

\noindent\textbf{{(Property 3)} Code is visually sparse:}
Indentation, blank lines, and short statements leave substantial unused space in rendered code, while line-preserving pagination further limits page density. Hence, the renderer must balance per-page density and page count rather than simply shrinking fonts.

To validate these effects on real repository code, we run a single-read rendering study over sampled source-code windows. We render windows of different sizes under each nominal ratio and measure the resulting prompt-token cost without introducing multi-turn agent interaction.

\begin{figure}[t]
    \centering
    \includegraphics[width=0.85\linewidth]{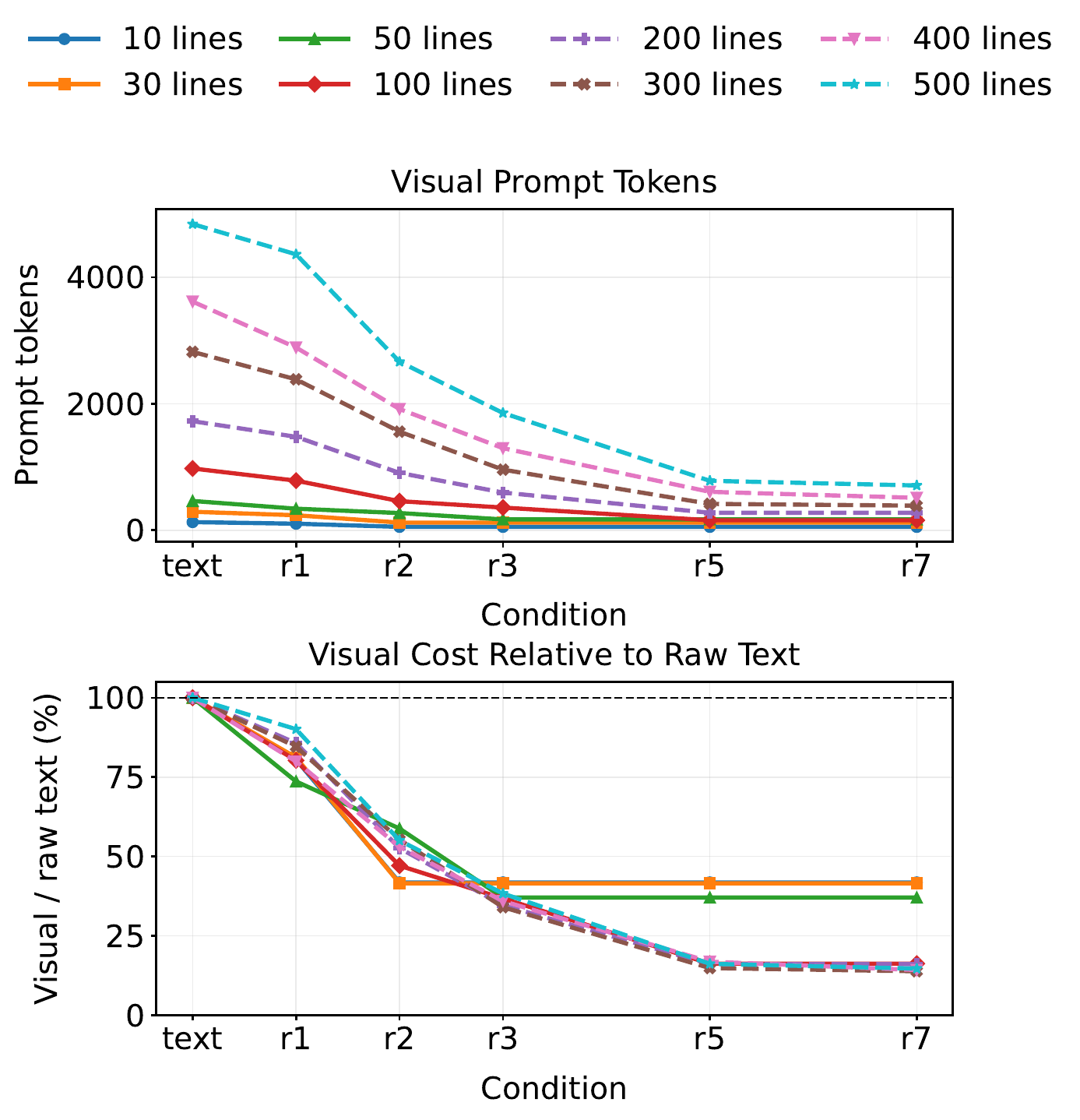}
    \caption{Single-read visual token cost across varying window sizes and nominal compression ratios. The two panels jointly show that token counts do not decrease linearly: small windows rapidly encounter an operational cost floor because they collapse to the same rendering configuration. Higher compression ratios become effective only for more extreme large-window settings, reducing cost to 15\%--40\% of raw text, but with diminishing marginal returns and inherent readability risks.}
    \label{fig:single-read-rendering}
    \vspace{-15pt}
\end{figure}

Figure~\ref{fig:single-read-rendering} shows that visual rendering already reduces cost at $r{=}1$, suggesting that modality conversion itself provides a denser representation than raw text tokenization.
However, increasing the nominal ratio does not produce smooth or proportional savings.
For small windows, the visual cost quickly reaches a floor: for example, 10-line windows cost 54.0 tokens for all $r{\geq}2$.
Clearer savings appear only for larger windows, but the need to preserve code structure and readability imposes a practical ceiling, leading to diminishing marginal returns.
We also observe that at higher $r$, rendered code can sometimes suffer from readability failures, introducing a practical trade-off beyond token cost.
The full token values and readability examples are provided in Appendix~\ref{app:rendering-examples}.

This matters for repository-level agents because they usually need small targeted regions rather than broad context. Large rendered windows may appear efficient under visual compression, but also include irrelevant code beyond the useful evidence.

\section{Controlled Experimental Design and Framework}
\label{sec:experimental-setup}

The attribution challenge in \S\ref{subsec:setup} and the non-smooth cost surface in \S\ref{subsec:cost-surface} make a single text-to-visual swap difficult to interpret. As shown in Figure~\ref{fig:intro}, we therefore evaluate visual code through a controlled progression of three agent configurations: the official mini SWE-agent baseline (mini), the Pre-Localized Bash Agent (PLB), and the Locator--Editor Tool Agent (LET). Below, we summarize the key design of each configuration (details are provided in Appendix~\ref{app:pipeline-details}).

\subsection{mini SWE-agent Baseline}
We use the official mini SWE-agent v2 \citep{yang2024sweagent} as our baseline, since it is designed to evaluate the model's coding ability in a fair, standard setting. It runs a single-stage bash-first ReAct \citep{yao2023react} loop with the full mix of repository exploration, code reading, editing, and test execution. 

\subsection{Shared Protocol for PLB and LET}

\paragraph{Evaluation protocol.}
All configurations share the same evaluation. We sample 100 instances from SWE-bench Verified under a fixed seed, stratified so that every repository represented in the full benchmark appears in the subset. All configurations use the same backbone $M_\theta$, rendering pipeline, official SWE-bench harness, and are evaluated under one text condition and four visual conditions indexed by $r \in \{1,3,5,7\}$. For each (instance, condition) pair, we run a single trajectory and report pass@1 correctness under the official SWE-bench evaluation (Eq.~\ref{eq:swebench-eval}), together with total prompt-token consumption over the full trajectory.

\paragraph{Repository indexing.}
Both PLB and LET share a static repository index. For each repository, we parse every Python source file with a tree-sitter parser \citep{treesitter} and extract each top-level and nested function, class, and method as a symbol $\sigma$, storing its file path, kind, qualified name, line span, parent, base classes, and call set. On top of these per-symbol records we construct a lightweight repository graph $G$ whose edges encode parent--member, file-import, inheritance, and call relations. The resulting static index is
\[
\mathcal{I}(\textsc{repo}) \;=\; \bigl(\Sigma,\ \mathcal{F},\ G,\ \mathcal{B}\bigr),
\]
where $\Sigma$ is the symbol set, $\mathcal{F}$ the directory tree, $G$ the lightweight graph, and $\mathcal{B}$ a sparse BM25 retriever over the chunk-text corpus $\{\mathrm{chunk}(\sigma) : \sigma \in \Sigma\}$. With the repository organized as $\mathcal{I}$, a textual query $q$ localizes repository context by returning the top-$k$ matched symbols together with their aligned chunk text,
\[
\mathcal{R}(q;\mathcal{I})
=
\bigl((\sigma_{(i)}, c_{(i)})\bigr)_{i=1}^{k},
c_{(i)}=\mathrm{chunk}(\sigma_{(i)}).
\]

\paragraph{Localize and Package operators.}
On top of $\mathcal{I}$ we define two procedures that are shared by both PLB and LET and treated as structured repository tools. Let $q$ denote a textual query, typically synthesized from the issue text $x$. We write $D_{\text{loc}}$ and $D_{\text{pkg}}$ for the concrete system-side implementations of the Localize and Package operators. \emph{Localize} performs coarse repository localization:
\[
\mathcal{L}(x, q)
\;=\;
D_{\text{loc}}\!\bigl(x,\ q,\ \mathcal{R}(q;\mathcal{I}),\ G\bigr).
\]
Operationally, \emph{Localize} combines retrieved symbol chunks with issue-aware matching and lightweight graph expansion to produce ranked candidate files and fault-relevant symbols. \emph{Package} then refines these candidates into a structured context package:
\[
\mathcal{P}(x, q)
\;=\;
D_{\text{pkg}}\!\bigl(x,\ q,\ \mathcal{L}(x,q),\ \mathcal{I}\bigr),
\]
which uses the coarse candidates implied by $\mathcal{L}(x,q)$ to retrieve finer-grained repository evidence and serialize it into a compact multi-view context package. PLB invokes $\mathcal{L}$ and $\mathcal{P}$ once offline to produce the pre-localization brief $\widetilde{\mathcal{L}}(x)$ and package $\hat{\mathcal{P}}(x)$, while LET exposes them as online tools for the locator. Implementation details, prompt templates, plan schemas, and visual-intervention boundaries are provided in Appendix~\ref{app:pipeline-details}.

\subsection{Pre-Localized Bash Agent (PLB)}

PLB keeps the same bash-first interaction style as mini, but prepends an offline pre-localization pass before the online repair loop. The motivation is to reduce the bias introduced by unguided exploration, which can strongly perturb long trajectories and is orthogonal to the modality effect we want to study. For each instance, we set $q := x$, retrieve $\mathcal{R}(x;\mathcal{I})$ from the shared repository index, and use two single-shot LLM calls to produce a compact starting package
\[
\widetilde{\mathcal{P}}(x)
\;=\;
\mathrm{E}_{\text{pkg}}\!\bigl(
M_\theta(\pi_{\text{pkg}}(x,\ \mathcal{R}(x;\mathcal{I}),\ \widetilde{\mathcal{L}}(x)))
\bigr),
\]
where $\widetilde{\mathcal{L}}(x)$ is an intermediate localization brief produced by an analogous call. The package is computed once per instance, shared across text and visual conditions, and provided as the agent's starting context. The agent then runs the same bash-first loop until \texttt{submit}; under visual conditions, source-code outputs from the model's code-reading commands are rendered as images, while non-code outputs remain in text form. PLB therefore evaluates visual rendering in a setting that remains close to the model's native bash-based workflow, while suppressing part of the trajectory variance caused by unconstrained repository exploration.

\subsection{Locator--Editor Tool Agent (LET)}
LET adopts a two-stage decoupled design that separates repository exploration from the trial-and-error editing stage. This allows us to study whether rendered raw code still adds value when the agent already accesses structured repository information.

  \noindent\textbf{Locator.}
  In the locator phase, the agent operates with tools
  \{\texttt{Localize},\ \texttt{Package},\ \texttt{view\_code},\ \texttt{commit\_plan}\}.
  Here \texttt{Localize} and \texttt{Package} correspond to the shared operators
  $\mathcal{L}$ and $\mathcal{P}$. In the locator phase, the agent iteratively localizes likely edit sites, packages candidate context, and inspects source windows via \texttt{view\_code}.

  \noindent\textbf{Handoff.}
The locator phase ends by emitting a committed plan $\Pi(x)$ with likely targets, a reproduction command, issue-relevant evidence, and referenced source code windows. The full locator transcript is then discarded.

  \noindent\textbf{Editor.}
  The editor starts from  plan $\Pi(x)$ and operates with 
\{\texttt{Localize}, \texttt{Package}, \texttt{view\_code}, \texttt{bash}, \texttt{submit}\}. 
It retains the repository-reading tools for evidence re-checking, while the main focus of this phase is editing and test execution through \texttt{bash}. 
We keep repair in bash based on our preliminary trials: adding separate editing tools introduced redundant interactions and consistently weakened end-to-end performance.

\section{Results and Analysis}
\begin{figure}[t!]
    \centering
    \includegraphics[width=1\linewidth]{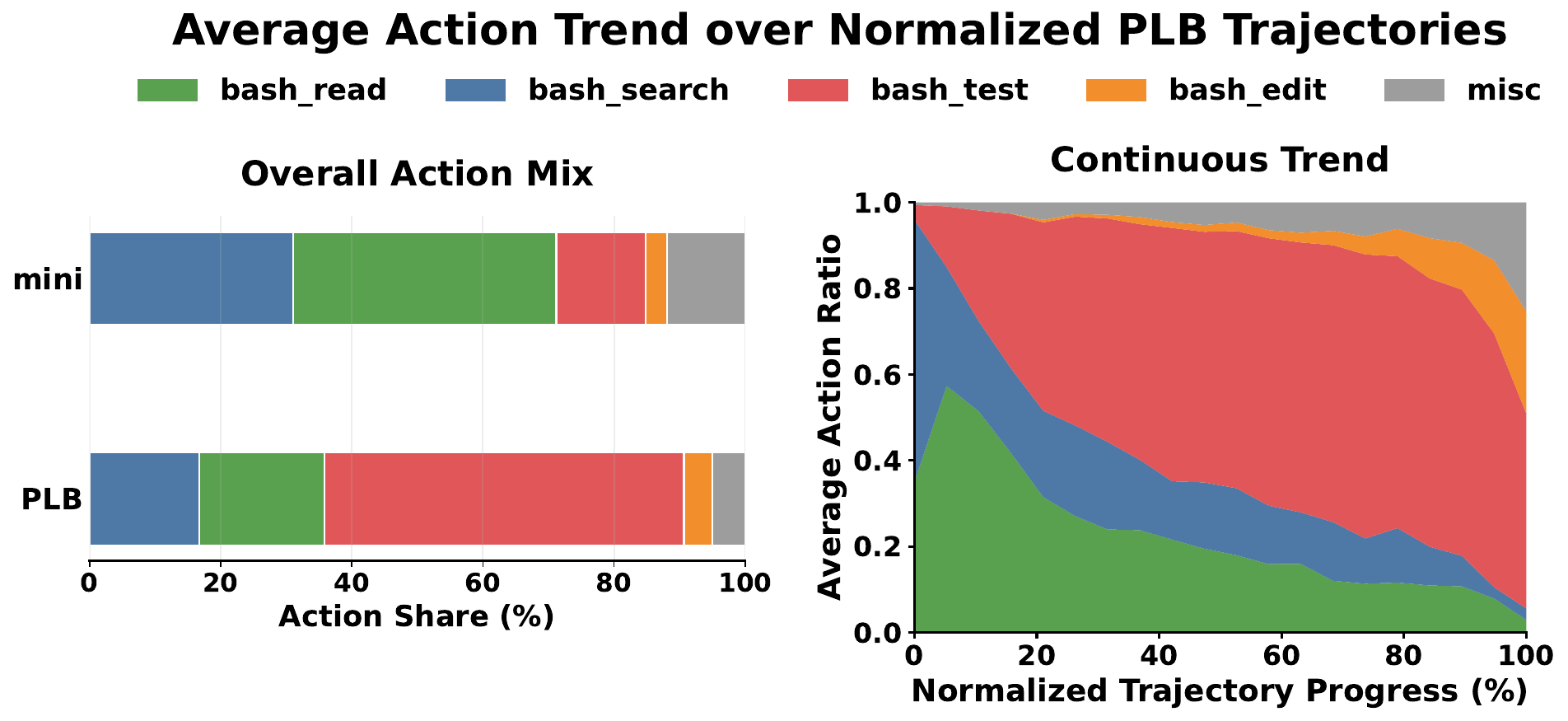}
    \caption{Behavioral shift from mini to PLB. Pre-localization reduces unguided repository exploration, with read actions declining over time.}
    \label{fig:plb-action-trend}
    \vspace{-10pt}
\end{figure}
\begin{figure}[t!]
    \centering
    \includegraphics[width=1\linewidth]{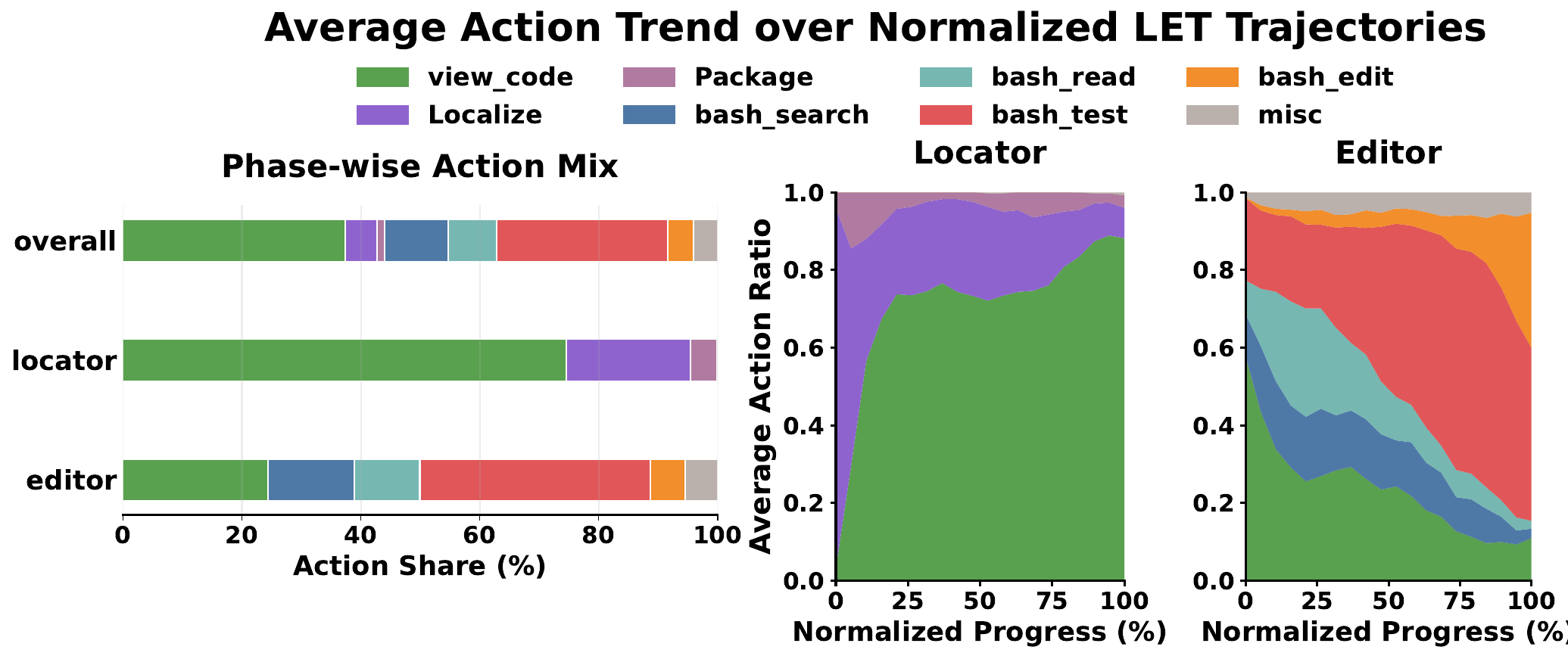}
    \caption{Behavioral decomposition of LET. The locator is dedicated to repository understanding, while the editor focuses on trial-and-error repair.}
    \label{fig:let-action-trend}
    \vspace{-15pt}
\end{figure}

\begin{table*}[t!]
    \centering
    \caption{Raw accuracy and average prompt-token cost on Qwen3.5. Tokens are in millions; token-gain columns report text reduction over mini and visual reduction over mini.}
    \label{tab:qwen35-raw}
    \setlength{\tabcolsep}{2.7pt}
    \renewcommand{\arraystretch}{1.02}
    \begin{tabular}{@{}llcccccc|cccccc|cc@{}}
      \toprule
      \multirow{2}{*}{agent}
      & \multirow{2}{*}{scope}
      & \multicolumn{6}{c|}{Accuracy (\%)}
      & \multicolumn{6}{c|}{Avg tokens (M)}
      & \multicolumn{2}{c}{Token reduction} \\
      &
      & mini & text & $r{=}1$ & $r{=}3$ & $r{=}5$ & $r{=}7$
      & mini & text & $r{=}1$ & $r{=}3$ & $r{=}5$ & $r{=}7$
      & text & visual range \\
      \midrule
      \multirow{2}{*}{PLB}
      & normal & 70.0 & 72.0 & \best{73.0} & 71.0 & 69.0 & 70.0
               & 4.13 & 2.91 & 2.43 & 1.72 & \best{1.47} & 1.53
               & $1.42{\times}$ & \gain{$1.70$--$2.81{\times}$} \\
      & wide   & 70.0 & 70.0 & \best{72.0} & 71.0 & 70.0 & 71.0
               & 4.13 & 3.13 & 2.88 & 2.27 & \best{1.95} & 2.03
               & $1.32{\times}$ & \gain{$1.43$--$2.12{\times}$} \\
      \midrule
      \multirow{2}{*}{LET}
      & normal & \best{70.0} & 69.0 & 68.0 & \best{70.0} & 65.0 & 64.0
               & 4.13 & 1.94 & 1.84 & \best{1.69} & 1.74 & 1.78
               & $2.13{\times}$ & \weakgain{$2.24$--$2.44{\times}$} \\
      & wide   & \best{70.0} & 67.0 & 69.0 & 69.0 & 68.0 & 66.0
               & 4.13 & 1.73 & 1.78 & 1.80 & 1.77 & \best{1.64}
               & $2.39{\times}$ & \weakgain{$2.29$--$2.52{\times}$} \\
      \bottomrule
    \end{tabular}
\end{table*}
  
\begin{table*}[h]
    \centering
    \caption{Raw accuracy and average prompt-token cost on Kimi-K2.5. Tokens are in millions; token-gain columns report text reduction over mini and visual reduction over mini.}
    \label{tab:kimi-raw}
    \setlength{\tabcolsep}{2.7pt}
    \renewcommand{\arraystretch}{1.02}
    \begin{tabular}{@{}llcccccc|cccccc|cc@{}}
      \toprule
      \multirow{2}{*}{agent}
      & \multirow{2}{*}{scope}
      & \multicolumn{6}{c|}{Accuracy (\%)}
      & \multicolumn{6}{c|}{Avg tokens (M)}
      & \multicolumn{2}{c}{Token reduction} \\
      &
      & mini & text & $r{=}1$ & $r{=}3$ & $r{=}5$ & $r{=}7$
      & mini & text & $r{=}1$ & $r{=}3$ & $r{=}5$ & $r{=}7$
      & text & visual \\
      \midrule
      \multirow{2}{*}{PLB}
      & normal & 69.0 & 69.0 & \best{70.0} & 69.0 & 65.0 & \best{70.0}
               & 2.68 & 2.27 & 2.13 & 1.44 & \best{1.04} & 1.17
               & $1.18{\times}$ & \gain{$1.26$--$2.58{\times}$} \\
      & wide   & 69.0 & 71.0 & \best{72.0} & 69.0 & 69.0 & 71.0
               & 2.68 & 2.38 & 2.30 & 2.27 & 2.08 & \best{1.97}
               & $1.12{\times}$ & \gain{$1.17$--$1.36{\times}$} \\
      \midrule
      \multirow{2}{*}{LET}
      & normal & 69.0 & 68.0 & 68.0 & \best{70.0} & 64.0 & 60.0
               & 2.68 & 1.14 & 1.21 & \best{1.01} & 1.03 & 1.14
               & $2.35{\times}$ & \weakgain{$2.21$--$2.65{\times}$} \\
      & wide   & \best{69.0} & 67.0 & \best{69.0} & 65.0 & 65.0 & 66.0
               & 2.68 & \best{1.09} & 1.40 & 1.38 & 1.47 & 1.10
               & $2.46{\times}$ & \weakgain{$1.82$--$2.44{\times}$} \\
      \bottomrule
    \end{tabular}
    \vspace{-5pt}
\end{table*}
  
\subsection{Initial Context Reshapes Repair Behavior}

Mini exhibits the most interleaved behavior throughout the trajectory, making visual effects hard to
attribute.
Figures~\ref{fig:mini-agent-actions}, \ref{fig:plb-action-trend}, and
\ref{fig:let-action-trend} isolate progressively cleaner behavioral regimes.

\noindent\textbf{Pre-localization compresses unguided exploration into the early trajectory, allowing later behavior to shift toward focused test--edit repair.}
Compared with Mini, PLB enters the main repair loop with an initial starting point, and therefore no longer needs to sustain broad exploration throughout the trajectory.
The action mix reflects this shift: in Mini, \texttt{read} and \texttt{search} account for 40.1\% and 31.0\% of actions, whereas in PLB, the corresponding \texttt{bash\_read} and \texttt{bash\_search} shares drop to 19.2\% and 16.7\%; meanwhile, \texttt{bash\_test} becomes dominant at 54.8\%.
The normalized trajectory further shows that PLB does not remove exploration, but concentrates source re-checking and localization confirmation in the early stage; later behavior shifts toward test feedback, patch editing, and iterative repair.

\noindent\textbf{A more precise starting point, does not fundamentally change the focused repair pattern.}
LET moves repository understanding into the locator stage, where the locator can
repeatedly call \texttt{Localize} and \texttt{Package} to narrow the candidate
context. However, once control passes to the editor, the behavior does not
further collapse into a simple edit--test routine merely because the preceding
locator has provided more precise candidates. Instead, the editor still follows
a PLB-like repair pattern, involving source re-checking, test feedback, and patch
iteration around the given context. Thus, LET shows that the locator changes the
editor's information boundary, but not its core repair dynamics. Final
resolution still depends on whether the editor can effectively use this
information within the focused repair loop.

\subsection{Visual Compression in Agentic Repair}
\label{subsec:performance}
We evaluate each agent under two inspection scopes.
In the \emph{normal} scope, the prompt provides a recommended repair path;
in the \emph{wide} scope, the prompt explicitly encourages the agent to inspect broader repository context and code windows.
All prompt templates are provided in Appendix~\ref{app:prompt-templates}.

\noindent \textbf{Accuracy is mostly preserved.}
Across both backbones, PLB and LET generally remain in the same accuracy band as mini.
Since each condition contains 100 tasks, single-digit percentage-point fluctuations should be interpreted as task-level variation rather than a monotonic effect of visual compression.
However, aggressive compression can weaken this stability: the main exception appears in the LET setting, where the lowest accuracy drops to 60.0\%; by contrast, PLB is more stable overall.
This difference is related to how the two pipelines organize information and structure the repair process, which we discuss further below.

\noindent \textbf{Visual compression is most beneficial when raw source reading remains a major cost.}
The PLB text condition already reduces part of the overhead by limiting unguided repository exploration.
However, within this cleaner repair loop, the agent's repository understanding still relies primarily on source-code reading, so rendered code can further reduce the reading cost.
The largest PLB reductions reach $2.81\times$ on Qwen3.5 and $2.58\times$ on Kimi-K2.5, both under the normal scope.
By contrast, the gains are smaller under the wide scope, because broader inspection introduces more context beyond what is needed as repair evidence, consistent with \S\ref{subsec:cost-surface}.
Moreover, the reduction is not monotonic in $r$, reflecting both the non-smooth rendering cost surface and repeated calls caused by reduced readability in some rendered code.

\noindent \textbf{When structured repository information already reduces source-reading demand, visual compression has smaller marginal gains.}
Compared with PLB, LET further reduces token usage, showing that structured localization and packaging already make repository navigation cheaper.
Relative to mini, its largest reductions reach $2.52\times$ on Qwen3.5 and $2.65\times$ on Kimi-K2.5.
However, because the LET text baseline is already compact, visualizing raw code brings smaller and less stable additional gains.
This suggests that visual compression is best suited to settings where raw source reading remains a major cost; once structured repository information reduces that demand, the remaining overhead comes more from editing, testing, and patch iteration.

\begin{table}[t]
  \centering
  \small
  \setlength{\tabcolsep}{5pt}
  \caption{LET cost breakdown on Kimi-K2.5. The top block reports the phase-level token split; the bottom block reports editor-cost buckets by locator evidence and final outcome.}
  \label{tab:let-cost-breakdown}
  \begin{tabular}{lcc}
    \toprule
    \multicolumn{3}{l}{\textbf{Phase-level token split}} \\
    \addlinespace[2pt]
    group & locator & editor \\
    \midrule
    all          & 0.19M / 23.7\% & 1.18M / 76.3\% \\
    resolved     & 0.18M / 28.0\% & 0.83M / 72.0\% \\
    non-resolved & 0.22M / 16.3\% & 1.78M / 83.7\% \\
    \addlinespace[5pt]
    \multicolumn{3}{l}{\textbf{Editor-cost buckets}} \\
    \addlinespace[2pt]
    bucket & share & avg editor tok. \\
    \midrule
    high evidence, $\checkmark$ & 54.5\% & 0.72M \\
    high evidence, $\times$     & 12.0\% & 1.77M \\
    mid evidence, $\checkmark$  & 4.0\%  & 1.15M \\
    mid evidence, $\times$      & 9.5\%  & 1.75M \\
    low evidence, $\checkmark$  & 4.8\%  & 1.84M \\
    low evidence, $\times$      & 15.2\% & 1.80M \\
    \bottomrule
  \end{tabular}
  \vspace{-15pt}
\end{table}

\subsection{Locator--Editor Contribution Analysis}
\noindent\textbf{Trial-and-Error Matters More Than Understanding the Repository.}
The two-stage LET decomposition further shows that more precise localization does not automatically translate into lower repair cost; it mainly changes the editor's starting point, rather than the downstream trial-and-error process itself.
We use Kimi-K2.5 as a representative case.
As shown in Table~\ref{tab:let-cost-breakdown}, the locator accounts for only a
small fraction of the total cost, while the editor consumes most of the tokens.
This imbalance becomes even stronger on non-resolved tasks, where the editor
accounts for 83.7\% of the total token cost.
Thus, once LET proposes a candidate repair direction, the dominant cost shifts
to the editor-side edit--test--verify loop.

\begin{figure}[t]
  \centering
  \includegraphics[width=0.8\linewidth]{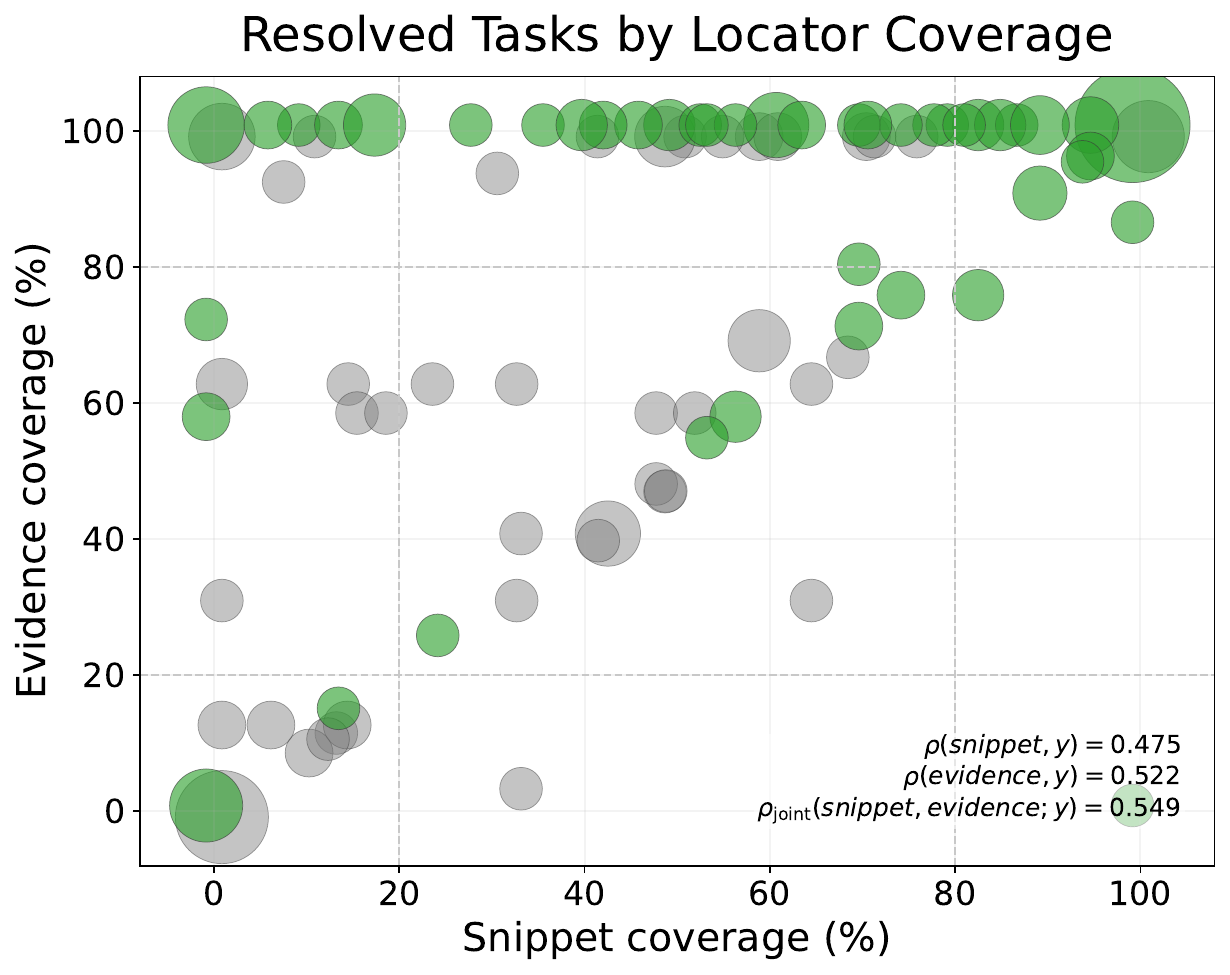}
  \caption{Task-level LET outcomes by locator coverage. Each bubble aggregates runs with the same snippet and evidence coverage; green denotes resolved outcomes and gray denotes non-resolved outcomes.}
  \label{fig:let-coverage-bubbles}
  \vspace{-15pt}
\end{figure}

We then relate locator coverage to editor cost.
For each LET task, we measure whether the locator commits patch-relevant
snippets and whether its observed source windows cover the gold-patch region.
Figure~\ref{fig:let-coverage-bubbles} shows that coverage is positively
correlated with resolution, but the relationship is not deterministic.
Better coverage improves the editor's initial condition, but does not guarantee
that the downstream patch will be correctly generated, tested, and refined.

The editor-cost buckets make this distinction clearer
(Table~\ref{tab:let-cost-breakdown}).
High-evidence tasks that resolve are the cheapest, requiring only 0.72M editor
tokens on average. However, once high-evidence tasks fail, their editor cost
still rises to a level comparable to other failures. Conversely, some
low-evidence tasks can still be rescued, but only through a more expensive
trial-and-error process. Therefore, localization mainly helps by increasing the chance that a task enters
the low-cost ``high-evidence and resolved'' regime. In difficult cases, the
remaining cost is driven by repeated verification, error correction, and patch
refinement within the patch--test loop.

\section{Conclusion}

We studied whether rendered code can serve as operational context for repository-level coding agents, beyond static code-understanding settings. On SWE-bench Verified, across Kimi-K2.5 and Qwen3.5, rendered code consistently reduces prompt-token cost while largely preserving end-to-end repair accuracy. However, the benefit is not a simple function of the nominal compression ratio: rendering cost is non-smooth, further compression does not always yield further token savings, and aggressive settings can make accuracy less stable.

Our controlled agent progression clarifies where this benefit comes from. PLB shows that rendered code is most useful when repository understanding still depends heavily on raw source-code reading, where visual representation can substantially reduce the cost of inspection. LET shows the complementary limit: once structured localization and packaging reduce the need for raw-code reading, much of the remaining cost shifts to editing, testing, and patch iteration. In this regime, visual compression still helps in some cases, but its marginal benefit becomes smaller and less reliable.

Overall, our results suggest that rendered code is a viable compression
mechanism for coding agents, but its benefit is conditional on whether raw
source reading remains a real bottleneck. For repository-level repair, the
central challenge is therefore not only how to compress code, but how to place
that compression inside the broader edit workflow.

\section*{Limitations}

\noindent\textbf{Evaluation Scope and Generalizability.}
While this exploratory study spans multiple multimodal backbones and a wide array of repository types, the findings reflect a controlled evaluation environment. Therefore, these initial insights warrant further validation in production-grade enterprise environments before extending these findings to definitive implementations of large-scale production solutions.


\noindent\textbf{Agent Abstractions and Production Orchestration.}
Our evaluation relies on PLB and LET as controlled, rigorous abstractions designed for clean attribution, rather than exhaustive production-grade configurations. While practical industrial orchestrations or multi-agent workflows may integrate advanced planning, retrieval, long-term memory, and collaborative verification, our decoupled framework serves as a vital blueprint. These initial insights establish foundational principles that help navigate where visual context can be most strategically deployed when transitioning to highly complex, enterprise-scale agentic systems.



\bibliography{custom}

\onecolumn
\newpage
\twocolumn
\appendix

\section{Action Type Categorization}
\label{appendix:action-types}

For Figure~\ref{fig:mini-agent-actions} we categorize each bash command emitted by mini SWE-agent v2 into one of five action types based on the command verb and its observed effect on the repository state:
\begin{itemize}
  \item \textbf{Read}: source-reading commands that return code-bearing stdout (e.g., \texttt{cat}, \texttt{sed -n}, \texttt{head}, \texttt{tail}, \texttt{nl -ba} with line slicing, and file viewers wrapped in similar idioms).
  \item \textbf{Search}: repository navigation and lookup commands that return non-code listings (e.g., \texttt{ls}, \texttt{find}, \texttt{grep}, \texttt{rg}, directory tree inspection).
  \item \textbf{Edit}: commands that mutate repository files (in-place editors, redirection into a file, patch application).
  \item \textbf{Test}: commands that execute the project test suite or a reproduction script (e.g., \texttt{pytest}, \texttt{python -m unittest}, project-local runners, repro scripts).
  \item \textbf{Misc}: remaining shell actions that do not fit the four categories above, including environment setup, dependency inspection, shell plumbing, and short diagnostic commands whose output
  is not classified as source reading, repository search, editing, or testing.
\end{itemize}
The temporal-trend panel of Figure~\ref{fig:mini-agent-actions} bins each trajectory step by its normalized position $t/T$ within that trajectory before aggregating across instances.

\section{Rendering Pipeline and Examples}
\label{app:rendering}

\subsection{Rendering Pipeline Details}
\label{app:rendering-details}
  \paragraph{Rendering objective.}
  Given a code window $x$ and a nominal compression ratio $r \ge 1$, the
  renderer maps $r$ to a target visual budget
  \[
  B(x,r) = \frac{\widehat T_{\text{text}}(x)}{r},
  \]
  where $\widehat T_{\text{text}}(x)$ is the renderer-side text-cost proxy. It
  then selects a rendering configuration $\theta=(\rho,f,p)$ by solving
  \[
  \theta^{*}(x,r)
  =
  \arg\min_{\theta\in\Theta}
  \mathcal{L}\!\bigl(R(x;\theta),\,B(x,r)\bigr),
  \]
  where $\mathcal{L}$ balances target-budget matching against layout efficiency
  and page overhead.

  \paragraph{Rendering settings.}
  All windows are rendered with a monospace font, line height $1.2$, page margin
  equal to $1\%$ of page width, syntax highlighting enabled, and source newlines
  preserved exactly. Tabs are expanded to four spaces, the language is inferred
  from the filename when needed, and pages are cut only at source-line
  boundaries so wrapped lines are not split across pages. To keep rendered code
  usable in the agent setting, we additionally apply lightweight readability
  safeguards during rendering, preventing layouts that are nominally cheap but
  visually too small to inspect reliably. In our agent setting, we also keep
  line numbers aligned with the rendered code window so that references passed
  between tools, plans, and source windows remain faithful to the underlying
  repository lines.

  \paragraph{Discrete search space.}
  The search space is explicitly discrete: resolution, font size, and page
  count are all drawn from finite candidate sets. Hence the renderer only
  realizes a discrete family of layouts, so nearby nominal budgets may collapse
  to the same configuration.

  \paragraph{Token term.}
  The token term matches the rendered cost to the nominal budget. Let
  $\widehat{T}_{\text{img}}(\theta)$ denote the renderer's internal visual-cost
  proxy. Then
  \[
  \mathcal{L}_{\text{token}}
  \propto
  \left|
  \frac{\widehat{T}_{\text{img}}(\theta)}{B(x,r)} - 1
  \right|.
  \]
  This term pulls the search toward configurations whose estimated image-token
  cost is close to the target implied by $r$.

  \paragraph{Layout term.}
  The layout term penalizes unreadable or inefficient code layouts. Define the
  fill ratio
  \[
  \phi(x,\theta)
  =
  \frac{\text{occupied area}}{\text{available text area}}.
  \]
  Then $\mathcal{L}_{\text{layout}}$ decreases as $\phi(x,\theta)$ approaches a
  readable operating range. This matters because code is visually sparse:
  indentation, blank lines, short lines, and line-preserving pagination all limit
  how densely the page can be packed. As a result, increasing $r$ does not
  translate into proportional savings.

\begin{table*}[t]
\centering
\small
\begin{tabular}{c|c|ccccc}
\toprule
Window & Text tokens & \multicolumn{5}{c}{Visual cost: tokens (pages)} \\
\cmidrule(lr){3-7}
Lines & Raw text & r1 & r2 & r3 & r5 & r7 \\
\midrule
10  & 129.2  & 104.0 (1.00)   & 54.0 (1.00)   & 54.0 (1.00)   & 54.0 (1.00)   & 54.0 (1.00) \\
30  & 292.8  & 237.7 (1.60)   & 121.5 (1.00)   & 121.5 (1.00)   & 121.5 (1.00)   & 121.5 (1.00) \\
50  & 463.3  & 341.1 (3.13)  & 272.3 (2.73)  & 171.5 (2.70)  & 171.5 (2.70)  & 171.5 (2.70) \\
100 & 977.0  & 784.0 (4.13)  & 460.0 (4.40)  & 358.9 (4.87)  & 158.2 (4.97)  & 158.2 (4.97) \\
200 & 1724.0 & 1477.3 (3.50)  & 906.8 (6.63)  & 596.4 (6.40)  & 276.5 (9.70)  & 276.5 (9.70) \\
300 & 2819.7 & 2385.1 (3.50)  & 1558.6 (5.63)  & 958.4 (8.70)  & 417.1 (11.10) & 389.0 (13.43)\\
400 & 3613.9 & 2888.2 (5.97)  & 1916.4 (5.10)  & 1297.8 (11.80) & 607.4 (8.60)  & 512.7 (17.10) \\
500 & 4839.6 & 4360.8 (6.27) & 2663.7 (7.47) & 1853.3 (7.83) & 781.6 (12.97) & 708.2 (14.17) \\
\bottomrule
\end{tabular}
\caption{Single-read rendering cost on SWE-bench Verified \cite{chowdhury2024swebenchverified} repository code. Each cell reports average prompt tokens, with the average number of rendered pages in parentheses.}
\label{tab:single-read-rendering}
\end{table*}

  \paragraph{Page term and realized cost.}
  The page term penalizes excessive fragmentation, favoring fewer pages and
  discouraging high-page-count configurations under small budgets. Even after a
  configuration is selected, the realized API-side cost
  \[
  T_{\text{api}}(R)
  =
  c_{0}
  +
  \sum_{j=1}^{p} T_{\text{img}}(I_j)
  \]
  need not equal the nominal target $B(x,r)$: the search is performed under an
  internal proxy, the candidate set is discrete, and page count adds overhead.
  Therefore $r$ should be read as a nominal renderer-side control variable
  rather than the realized prompt-cost ratio.

\subsection{Rendering Examples and Token Values}
\label{app:rendering-examples}

\paragraph{Single-read rendering cost.}
Table~\ref{tab:single-read-rendering} reports the standalone token cost of
rendering SWE-bench Verified repository code across different window sizes and
nominal compression ratios. Each cell gives the average prompt tokens, with the
average number of rendered pages in parentheses. The table provides the raw
values behind the cost-surface discussion, showing that small windows quickly
hit a visual cost floor, while larger windows obtain stronger but non-smooth and
bounded savings.

\begin{figure}[h]
    \centering
    \includegraphics[width=1\linewidth]{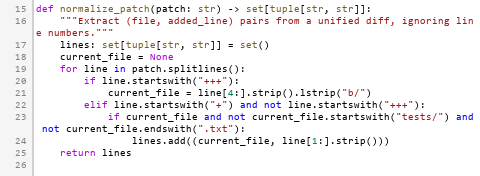}
    \caption{Rendered visual counterpart of the raw-text code excerpt in Listing~\ref{raw_code}. The renderer preserves the source layout and line-number alignment when converting the code window into an image representation.}
    \label{fig:rendered-code-example}
\end{figure}

\begin{figure}[h]
    \centering
    \includegraphics[width=1\linewidth]{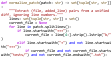}
    \caption{Example readability failure under a higher visual compression ratio. Although the rendered page remains compact, dense layout and small text can make source code harder to inspect reliably.}
    \label{fig:rendering-readability-failure}
\end{figure}

\paragraph{Illustrative examples.}
\label{app:rendering-readability}
Listing~\ref{raw_code} and Figure~\ref{fig:rendered-code-example} show a typical text-to-visual conversion used in our experiments. The renderer converts the raw source excerpt into an image while preserving the original line structure and line-number alignment, clarifying the form of code-bearing observations seen by the agent under visual conditions. Figure~\ref{fig:rendering-readability-failure} further illustrates a failure mode at higher visual compression ratios: increasing page density can make the rendered code noticeably harder to read, creating a practical readability trade-off beyond token cost.

\begin{lstlisting}[
  style=promptstyle,
  label={raw_code},
  caption={Raw-text source excerpt used for the rendering example.}
]
def normalize_patch(patch: str) -> set[tuple[str, str]]:\n    """Extract (file, added_line) pairs from a unified diff, ignoring line numbers."""\n    lines: set[tuple[str, str]] = set()\n    current_file = None\n    for line in patch.splitlines():\n        if line.startswith("+++"):\n            current_file = line[4:].strip().lstrip("b/")\n        elif line.startswith("+") and not line.startswith("+++"):\n            if current_file and not current_file.startswith("tests/") and not current_file.endswith(".txt"):\n                lines.add((current_file, line[1:].strip()))\n    return lines
\end{lstlisting}

\section{Pipeline Details}
\label{app:pipeline-details}

\subsection{Localize and Package Operators}
\label{app:loc-pkg}

The Localize operator is defined as
\[
\mathcal{L}(x,q)
=
D_{\text{loc}}\!\left(x,\ q,\ \mathcal{R}(q;\mathcal{I}),\ G\right),
\]
and the Package operator is defined as
\[
\mathcal{P}(x,q)
=
D_{\text{pkg}}\!\left(x,\ q,\ \mathcal{L}(x,q),\ \mathcal{I}\right).
\]
Both operators are introduced in \S\ref{sec:experimental-setup}; this appendix provides the implementation details omitted from the main text.

\paragraph{Inside $D_{\text{loc}}$.}
Given the retrieved symbol-chunk pairs $\mathcal{R}(q;\mathcal{I})$, 
$D_{\text{loc}}$ runs three sub-procedures:
\begin{itemize}
    \item \textbf{Issue-aware identifier resolution:} extracts likely function, class, and module names from the issue text and matches them against $\Sigma$.
    \item \textbf{Traceback and file-line hint resolution:} parses any traceback or explicit file/line reference in the issue and resolves it to the enclosing symbol.
    \item \textbf{Graph expansion:} expands over $G$ by adding parent symbols, file-import neighbors, base classes, and direct callers/callees of each matched symbol.
\end{itemize}
The resulting candidates are merged and ranked to produce the localization result returned by $\mathcal{L}(x,q)$.
\paragraph{Inside $D_{\text{pkg}}$.}
Given $\mathcal{L}(x,q)$, $D_{\text{pkg}}$ selects candidate files and optional focus symbols, then serializes a compact multi-view context from $\mathcal{I}$:
\begin{itemize}
    \item \textbf{File summaries:} provide each selected file's directory location and import-neighborhood summary.
    \item \textbf{Refined code windows:} extract source windows around each focus symbol, sliced by line span and lightly trimmed.
    \item \textbf{Search hints:} list identifiers and traceback anchors that the downstream agent may want to grep for.
    \item \textbf{Optional test pointers:} list tests that import or reference selected symbols.
\end{itemize}
The bundle is returned in a fixed JSON-like schema, so the PLB pre-localization pass and the LET locator tool consume it through the same parser.

\subsection{Pre-Localized Bash Agent Details}
\label{app:plb-details}

PLB consists of an offline pre-localization followed by a single-stage bash-first main agent.

\paragraph{Offline pre-localization.}
For each instance we construct $\mathcal{I}$ and set the retrieval query to the issue text:
\[
q := x.
\]
We then retrieve $\mathcal{R}(x;\mathcal{I})$, the top-$k$ symbol-chunk matches under BM25, and make two single-shot LLM calls. Let $\pi_{\text{loc}}, \pi_{\text{pkg}}$ denote fixed prompt templates and $\mathrm{E}_{\text{loc}}, \mathrm{E}_{\text{pkg}}$ denote structured JSON extraction operators. The first call produces a compact localization brief
\[
\widetilde{\mathcal{L}}(x)
\;=\;
\mathrm{E}_{\text{loc}}\!\bigl(\,
M_\theta(\pi_{\text{loc}}(x,\ \mathcal{R}(x;\mathcal{I})))
\,\bigr),
\]
and the second turns the same retrieved evidence together with $\widetilde{\mathcal{L}}(x)$ into the shared starting package
\[
\hat{\mathcal{P}}(x)
\;=\;
\mathrm{E}_{\text{pkg}}\!\bigl(\,
M_\theta(\pi_{\text{pkg}}(x,\ \mathcal{R}(x;\mathcal{I}),\ \widetilde{\mathcal{L}}(x)))
\,\bigr).
\]
The package contains a compact starting brief with refined candidate files and symbols, a reproduction hint, search hints, optional test pointers, and short notes. It is computed once per instance and shared across all PLB conditions.

\paragraph{Main agent interface.}
The downstream agent uses the tool set
\[
\mathcal{T}_{\text{bash}} = \{\texttt{bash},\ \texttt{submit}\}.
\]
It is initialized with $(x, \hat{\mathcal{P}})$ and runs a standard bash-first ReAct-style loop until \texttt{submit}.

\paragraph{Code observations.}
In PLB, \texttt{bash} is the only window onto repository source code. Under text conditions, code-bearing stdout is returned directly as text. Under visual conditions, when the issued command is recognized as a source-reading command and its stdout is interpreted as a code window, that window is rendered with CodeOCR at compression ratio $r$. Typical source-reading commands include \texttt{cat}, \texttt{sed -n}, \texttt{head}, \texttt{tail}, and \texttt{nl -ba} combined with line slicing. Under wide conditions, the detected code window may be expanded before it is returned or rendered. Non-code stdout remains in text form.

\subsection{Locator--Editor Tool Agent Details}
\label{app:let-details}

LET decomposes the repair process into a locator phase and an editor phase.

\paragraph{Locator interface.}
The locator operates with tool set
\[
\begin{aligned}
\mathcal{T}_{\text{loc}}
&=
\{\texttt{localize\_issue},\ \texttt{pack\_context},\\
&\quad \texttt{view\_code},\ \texttt{commit\_plan}\}.
\end{aligned}
\]

\paragraph{Localization tool.}
At trajectory step $t$, the agent synthesizes a short code-anchored query $q_t$ from the issue and its current hypothesis, and supplies it to the localization tool. The resulting operator is
\[
\mathcal{L}_t(x)
\;=\;
D_{\text{loc}}\!\bigl(x,\ q_t,\ \mathcal{R}(q_t;\mathcal{I}),\ G\bigr),
\]
i.e., the same Localize procedure from Appendix~\ref{app:loc-pkg} but driven by an online query rather than the static issue text.

\paragraph{Packaging tool.}
Unlike localization, packaging is not driven by a free-form textual query. After inspecting $\mathcal{L}_t(x)$, the agent explicitly chooses a small set of candidate file paths and optional focus symbols. The operator is then
\[
\mathcal{P}_t(x)
\;=\;
D_{\text{pkg}}\!\bigl(x,\ q_t,\ \mathcal{L}_t(x),\ \mathcal{I}\bigr),
\]
serialized as a compact multi-view structured context around the chosen candidates, including file/symbol context, refined code windows, and optional neighboring evidence.

\paragraph{Direct code inspection.}
The locator may inspect code directly through \texttt{view\_code}, which returns a raw source window specified by file path and line span. In visual conditions, locator-side code-bearing observations are rendered at compression ratio $r$; in text conditions they are returned as plain text.

\paragraph{Committed plan and handoff.}
The locator phase ends when the agent calls \texttt{commit\_plan}, which emits a structured plan
\[
\begin{aligned}
\Pi(x) = \bigl(\, & \texttt{target\_files},\ \\
                  & \texttt{target\_symbols},\ \\
                  & \texttt{reproduction\_command},\ \\
                  & \texttt{relevant\_snippets},\ \\
                  & \texttt{propagation\_targets},\ \\
                  & \texttt{notes},\ \\
                  & \texttt{evidence\_refs} \,\bigr).
\end{aligned}
\]
The locator transcript itself is dropped at the phase boundary. Only $\Pi(x)$ and the source windows referenced by \texttt{evidence\_refs} cross over into the editor's initial context.

\paragraph{Editor interface.}
The editor inherits the locator-side tools and adds shell access:
\[
\begin{aligned}
\mathcal{T}_{\text{ed}}
&=
\{\texttt{localize\_issue},\ \texttt{pack\_context},\\
&\quad \texttt{view\_code},\ \texttt{bash},\ \texttt{submit}\}.
\end{aligned}
\]
Edits are applied through \texttt{bash}, and the agent may re-invoke localization or packaging if the locator's plan turns out to be insufficient.

\paragraph{Visual intervention boundary.}
In the current implementation, the visual rendering intervention is applied on the locator side only. The editor re-reads source in text form even under visual LET conditions. Thus LET evaluates whether visual code still provides marginal value once repository exploration has already been enriched by structured online tooling.

\section{Reproducibility Details}
\label{app:reproducibility}

\paragraph{Backbones and API settings.}
All main experiments use two multimodal backbones: Kimi-K2.5 and Qwen3.5
(397B variant). For the
online coding agents, the model client uses temperature $0.0$, maximum output
length $4096$. The offline PLB pre-localization pass
uses the same endpoint with temperature $0.0$, maximum output length $1600$.

\paragraph{Task sampling and limits.}
The reported study focuses on Python repair tasks and uses a 100-instance stratified subset of SWE-bench Verified, the 500-instance human-validated subset designed to filter for clearer issue descriptions and problems resolvable from the provided repository context.
The subset is sampled with a deterministic seed so that every Python repository present in SWE-bench Verified appears in our evaluation.
The default global step limit is $250$.
In LET, the online phase budgets are further split into a locator budget of $40$ steps and an editor budget of $120$ steps.
All conditions share the same official SWE-bench harness and evaluation rule.

\paragraph{Condition families.}
Each backbone is evaluated under one text condition and four visual conditions
indexed by $r \in \{1,3,5,7\}$. For PLB and LET we also distinguish
\emph{normal} and \emph{wide} views. In the \emph{normal} setting, a code-reading request returns only the source span selected by the tool. 
In the \emph{wide} setting, the same request is expanded to include a larger surrounding window, so the agent receives more neighboring code before the observation is returned or rendered.

\section{Trajectory-Level Failure Case Studies}
\label{app:failure-case-studies}

Because LET separates the locator and editor trajectories, we select three cases from LET runs to make these trajectory-level effects easier to inspect. We include three representative cases. The goal is not to showcase best-looking examples, but to isolate distinct failure modes at the trajectory level.

\subsection{Case 1: Correct Localization, Slow or Incomplete Editing}

\paragraph{Task.}
\texttt{matplotlib\_\_matplotlib-26342} requires adding
\texttt{ContourSet.set\_paths} in
\texttt{lib/matplotlib/contour.py}, with the corresponding contour tests as the
gold verification target.

\paragraph{Comparison.}
We compare two experimental conditions, \texttt{LET-normal-visual-r7} and
\texttt{LET-wide-visual-r7}. Both localize the same relevant files,
\texttt{contour.py} and \texttt{test\_contour.py}, but diverge in the editor
stage. The normal visual run fails with \texttt{empty\_patch} despite 100\%
evidence coverage, whereas the wide visual run resolves the task with lower
evidence coverage (57.1\%) but a more focused patch-and-test loop.

\paragraph{Trajectory behavior.}
In the failing run, the locator already inspects the correct code region:
\texttt{ContourSet.get\_paths}, nearby collection methods, and the target test
file. The editor then repeatedly re-reads local windows
(\texttt{view\_code(contour.py, 1460--1480)},
\texttt{view\_code(collections.py, 35--220)},
\texttt{view\_code(contour.py, 930--1000)}) and issues grep-style checks, but
never commits to a concrete patch or a real verification command. The run
spends 121 editor steps and 2.75M editor tokens before terminating with no
patch. In contrast, the successful run moves into a direct edit loop: after a
small number of reads, it writes a temporary patch script, inserts
\texttt{set\_paths()}, runs a local smoke script, and then switches to focused
\texttt{pytest} commands over \texttt{test\_contour.py}.

\paragraph{Interpretation.}
This is the cleanest example of a trajectory where localization is already good
enough, but the editor does not convert that signal into a patch. The failure
mode is hesitation and repeated local rereading, not repository miss.

\subsection{Case 2: Locator Miss, Editor Rescue via Bash Search}

\paragraph{Task.}
\texttt{django\_\_django-11299} requires modifying
\texttt{django/db/models/sql/query.py}, specifically the logic around
\texttt{Query.\_add\_q}, so that SQL generated for a \texttt{CheckConstraint}
does not propagate incorrect aliasing behavior.

\paragraph{Comparison.}
We compare the visual and text conditions in LET at $r{=}1$. In the visual run,
the locator anchors the plan on \texttt{django/db/models/constraints.py} and
gets zero gold coverage. The text run, by contrast, directly surfaces
\texttt{query.py} with full evidence and snippet coverage. Despite this locator
miss, the visual run still resolves the task.

\paragraph{Trajectory behavior.}
The rescue happens in the editor, not through another structured localization
step. The editor falls back to bash-native search, using commands such as
\texttt{grep} over \texttt{query.py} and \texttt{django/db/models/} to find
relevant terms including \texttt{build\_where}, \texttt{with\_col\_aliases},
and \texttt{simple\_col}. It then re-reads the target region with
\texttt{view\_code}, writes a temporary reproduction script for the SQL
generation bug, edits \texttt{query.py}, and verifies the patch with
\texttt{python tests/runtests.py constraints}. This rescued visual run is much
more expensive than the clean text-guided run, taking 62 editor steps and
1.20M editor tokens versus 21 steps and 0.28M.

\paragraph{Interpretation.}
This case shows that editor rescue is possible after a locator miss, but it is
costly. The recovery comes from local bash search and rereading rather than
from re-invoking the structured locator, reinforcing that failures in
repository understanding can be shifted downstream into a more expensive
patch-search loop.

\subsection{Case 3: Correct Target, Incorrect Semantic Fix}

\paragraph{Task.}
\texttt{pydata\_\_xarray-6461} concerns
\texttt{xarray.where(..., keep\_attrs=True)} when the second argument is a
scalar. The gold patch is in \texttt{xarray/core/computation.py}, but the fix
requires preserving the intended \texttt{keep\_attrs} semantics across scalar
and variable inputs.

\paragraph{Comparison.}
We compare the text and visual LET conditions, with the visual run at $r{=}3$.
Both localize the correct file with 100\% evidence coverage, but only the
visual run resolves the task.

\paragraph{Trajectory behavior.}
The unresolved text run reproduces the error with a local smoke command and
then tests several nearby variants. It eventually patches
\texttt{keep\_attrs = lambda attrs, context: attrs[1]} as
\texttt{attrs[1] if len(attrs) > 1 else \{\}}. This avoids the index error, but
uses the wrong fallback semantics and fails the official test. The resolved
visual run binds the target test earlier, re-reads the relevant
\texttt{where()} implementation and test window, and converges to the stronger
fallback \texttt{attrs[1] if len(attrs) > 1 else attrs[0] if attrs else \{\}}.

\paragraph{Interpretation.}
This case is not about finding the right file or function. Both runs reach the
correct target, but one fails to recover the API-level semantics needed for the
patch. The bottleneck is therefore semantic repair inside the editor, not
repository understanding.

\section{Prompt Templates}
  \label{app:prompt-templates}

  For reproducibility, we include the raw prompt templates used by the released
  implementation rather than paraphrased summaries. The prompts are organized as
  paired \texttt{system} and \texttt{instance} templates. In our implementation,
  the \texttt{system} template defines persistent behavioral constraints, tool
  boundaries, and phase-specific responsibilities, while the \texttt{instance}
  template injects task-specific content such as the issue title, issue
  description, working directory, and locator-to-editor handoff context.

  We expose the core prompt templates for the three main agent roles in the
  paper: PLB, the LET locator, and the LET editor. These prompt pairs differ
  mainly in behavioral control rather than task content. The PLB prompt treats
  the pre-localized package as a starting hypothesis for a bash-native repair
  loop. The LET locator prompt forbids editing and testing, and instead requires
  the model to commit a structured plan. The LET editor prompt starts from that
  committed plan and encourages focused re-reading and repository-native
  verification before submission.

  The \emph{wide} setting is not implemented uniformly across pipelines. In PLB,
  the difference between normal and wide mainly comes from source-window
  expansion and rendering policy rather than from a distinct prompt template. In
  LET, however, the wide locator condition uses a dedicated prompt pair that
  explicitly encourages broader code inspection before plan commitment. We
  therefore include the LET wide-locator prompts below as the representative
  prompt-level intervention for broader repository exploration. The offline PLB
  pre-localization prompts are implemented directly in
  \texttt{experiments/prepackage\_runner.py}, where the system instruction and JSON
  schema for the Localize/Package calls are constructed in code.

  \paragraph{PLB prompt pair.}
  PLB keeps the agent in a bash-first repair loop. The system prompt defines the
  global workflow and boundaries, while the instance prompt injects the issue and
  the pre-localized package as the agent's starting context.

\begin{lstlisting}[
    style=promptstyle,
    float=*,
    floatplacement=t,
    label={prompt:plb-system},
    caption={PLB system prompt template.}
  ]
You are a single-stage repository coding agent.

You have exactly two tools:

1. `bash(command)`
   Use bash for search, precise source reading, focused probes, edits, repo-native tests,
   and final `git diff/status`.

2. `submit()`
   Submit exactly once after final diff inspection and at least one relevant
   repo-native test after the latest edit.

General rules:
- Work shell-first, similar to a mini-style coding agent.
- A lightweight prepackage may be provided as an additional user message. Treat it as a prior, not a proof.
- Verify code locations with bash before editing.
- Prefer focused reads such as `sed -n`, `head`, `tail`, `cat <small file>`, or `nl -ba ... | sed -n`.
- Do not `cat` large files blindly.
- Prefer one coordinated edit over many tiny rewrites.
- Modify source files only. Do not edit tests, docs, release notes, or benchmark metadata.
- Temporary scratch scripts are allowed for narrow semantic probes, but they do not replace a final repo-native test and must not be included in the submitted patch.
- If a relevant test passes and `git diff` only touches the intended source files, inspect final `git diff/status` and submit instead of continuing to browse.
- Do not repeat the same test command unless the code changed or you have a new hypothesis to check.

Visual-condition note:
- In visual runs, code-reading bash outputs may be returned as rendered images with short metadata.
- Treat those images as the exact code block to reason over.
\end{lstlisting}

\begin{lstlisting}[
    style=promptstyle,
    float=*,
    floatplacement=t,
    label={prompt:plb-instance},
    caption={PLB instance prompt template.}
  ]
<issue_title>
{{ task_title }}
</issue_title>
{% if task_description %}
<issue_description>
{{ task_description }}
</issue_description>
{% endif %}

<instructions>
You are solving a repository issue in `{{ cwd }}`.

Workflow:
1. Read the optional prepackage message if one is provided.
2. Use `bash` to confirm the likely files and symbols.
3. Use `bash` for exact source reading, focused edits, repo-native tests, and final `git diff/status`.
4. Before the first edit, write a short plan listing the files/symbols you expect to touch.
5. Run at least one relevant repo-native real test after your latest edit.
6. Finish with `submit()`.

Boundaries:
- Modify source files only.
- Do not edit tests.
- Do not edit docs or release notes.
- Do not emulate submit with bash.
- Keep source reads narrow unless a larger contiguous block is clearly justified.
- The prepackage is only a starting hypothesis. If bash evidence disagrees, trust the repo.
- Scratch scripts must stay temporary and must not be included in the patch.

Verification guidance:
- Django: `python tests/runtests.py <module>` from `/testbed`
- sympy: `bin/test <path>`
- Otherwise use the repo-native runner; `python -m pytest <path>` is a fallback when appropriate.

In each response:
- Briefly state what stage you are in.
- Make at least one tool call.
</instructions>
\end{lstlisting}

  \paragraph{LET locator prompt pair.}
  The LET locator prompt pair isolates repository understanding from downstream
  editing. The system prompt fixes the locator role and its tool boundaries,
  while the instance prompt instructs the agent to browse, inspect, and finally
  commit a structured plan without editing or testing.

\begin{lstlisting}[
    style=promptstyle,
    float=*,
    floatplacement=t,
    label={prompt:let-locator-system},
    caption={LET locator system prompt template.}
  ]
You are the LOCATOR phase of a two-phase repository coding agent.

Your job: identify exactly what needs to change in the repo. You will not edit
or test code yourself. When you are done, you call `commit_plan(...)` with a
structured plan, and the editor phase takes over with a fresh message history
seeded only by your plan.

Your step budget is small: aim to call `commit_plan(...)` once you have enough
code evidence to describe the edit site and the verification command. This is
a browse-style workflow: use coarse search to find promising locations, then
open precise details before committing the plan. Do not loop on broad query
rewrites once you have concrete file or symbol anchors to inspect. Once the
likely edit site, propagation targets, and verification command are clear,
stop browsing and commit the plan.

Available tools:
1. `localize_issue(issue, ...)`
   Coarse repository search. Synthesize a short code-anchored query from
   `<issue_title>` and `<issue_description>` (setting names, class names,
   function names). Pass that query as `issue`. Avoid generic verbs and
   natural-language paraphrases.
2. `pack_context(file_paths=[...], focus_symbols=[...])`
   Coarse browse/context packing. Use this when you want a compact view of
   likely files, related symbols, or structure before choosing exact lines.
   You may call it whenever a coarse context pack is more useful than another
   exact line window.
3. `view_code(file_path, start_line, end_line)`
   Fine-grained detail view. Use this to inspect exact current lines after
   coarse search. Prefer 100-250 lines per call so you see surrounding context
   (callers, related methods, type defs). Visual mode renders the upper end
   efficiently (~7 visual tokens per source line); text mode is more
   bandwidth-sensitive, stay near 100-150. Hard cap is 400 lines per call -
   subdivide larger reads. Do not widen reads after the edit path is already
   clear; the editor can work from the anchored regions you cite.
4. `commit_plan(target_files, target_symbols, reproduction_command, relevant_snippets, propagation_targets, notes)`
   Finalize. Call this exactly once when you are confident, and you should
   become confident quickly. If you have concrete file or symbol anchors, move
   to `pack_context(...)` and `view_code(...)` instead of continuing broad
   localization. Once the plan is specific enough for the editor to start,
   commit instead of gathering extra context.
You do NOT have `bash`, you do NOT have edit tools, and you do NOT have
`submit`. The editor phase will do the editing and run repo-native tests.

Plan rules:
- `target_files`: every source file the editor will read or modify.
- `target_symbols`: qualified names (e.g. `BaseConstraint.__init__`).
- `reproduction_command`: the editor's final repo-native verification command.
  Prefer `python tests/runtests.py <module>` for Django, `bin/test <path>`
  for sympy. Either form is recognized.
- `relevant_snippets`: the line ranges the editor must re-read before
  changing anything. The editor does not see your `view_code` history.
- `propagation_targets`: for each affected class or module, list the methods
  that must be touched together.  [Detailed generic Python state-propagation
  guidance is omitted here for readability.]
- `notes`: short free-form description of the intended semantic change.

General rules:
- Keep reasoning short and action-oriented.
- Do not paraphrase the issue. Use the exact code anchors.
- Never call `commit_plan` more than once.
- If a tool result is enough to write the plan, write the plan instead of
  calling another exploration tool.
\end{lstlisting}

\begin{lstlisting}[
    style=promptstyle,
    float=*,
    floatplacement=t,
    label={prompt:let-locator-instance},
    caption={LET locator instance prompt template.}
  ]
<issue_title>
{{ task_title }}
</issue_title>
{% if task_description %}
<issue_description>
{{ task_description }}
</issue_description>
{% endif %}

<instructions>
You are the LOCATOR phase of a two-phase agent solving a repository issue in
`{{ cwd }}`. You will not edit or test code. Your output is a single call to
`commit_plan(...)`.

Step budget: aim to commit the plan once you have enough code evidence to
describe the edit site, affected symbols, and a verification command. Use a
browse-style strategy, but choose the tool order yourself:
- `localize_issue(...)` is coarse search. Use a short code-anchored query
  synthesized from `<issue_title>` and `<issue_description>`. If the returned
  candidates miss the code path, refine the query with stronger file, format,
  class, function, or traceback anchors.
- `pack_context(...)` is coarse browsing/context packing. Use it when a compact
  multi-file view or symbol outline is more useful than exact line reading.
- `view_code(...)` is the fine-grained source view. Use it to inspect the
  relevant edit region with enough surrounding context, then stop widening the
  search once the repair target is clear and prepare `commit_plan(...)`.

You may use these three tools in any order and as many times as needed within
the step budget. Once the detail evidence is enough, call `commit_plan(...)`
exactly once. If the likely edit site and verification command are already
clear, do not keep comparing extra candidates.

Evidence handling:
  [Detailed evidence-filtering are omitted here for readability.]

Plan content:
- `target_files`: every source file the editor will read or modify.
- `target_symbols`: qualified names of every symbol that will change.
- `reproduction_command`: the editor's final repo-native verification command
  (`python tests/runtests.py <module>` or `bin/test <path>`).
- `relevant_snippets`: pinned `(file, start_line, end_line)` regions the
  editor must re-read up front.
- `propagation_targets`: for each class or module that gains new state,
  enumerate the sibling methods that may need to be updated together -
  typically `__init__`, `__eq__`, `__repr__`, `deconstruct`, `clone`, and
  `super()` call sites. Skip a method only when you can articulate why the
  new state must not appear there.
- `notes`: short description of the intended semantic change.
- `evidence_refs`: the Evidence IDs for source windows the editor should see
  up front.

Boundaries:
- You have no `bash`, no edit tools, no `submit`. The editor will do all
  modification and verification.
- Do not paraphrase the issue title or description before
  `localize_issue(...)`. Use the exact code anchors.
- Do not commit the plan before at least one `view_code` confirmed an edit
  region.

In each response:
- Briefly state what stage you are in (locating / packing / inspecting /
  committing).
- Make at least one tool call.
</instructions>
\end{lstlisting}

  \paragraph{LET editor prompt pair.}
  The LET editor prompt pair starts from the committed locator plan. The system
  prompt keeps the agent in an editor role with access to repository-native
  repair tools, while the instance prompt frames the locator plan as the first
  hypothesis and encourages focused verification before submission.

\begin{lstlisting}[
    style=promptstyle,
    float=*,
    floatplacement=t,
    label={prompt:let-editor-system},
    caption={LET editor system prompt template.}
  ]
You are the EDITOR phase of a two-phase repository coding agent. The locator
phase has already produced a structured plan. Treat that plan as a strong
starting point, not as an authority: execute it when it is coherent, but use
the available repo-reading tools to check nearby code, tests, and call paths
when the issue or the patch semantics require it.

Available tools:

1. `localize_issue(issue, ...)`
   Optional. Use when the locator plan looks incomplete, points at tests
   rather than source, or your edit/test results suggest another file or
   symbol is involved.

2. `pack_context(file_paths=[...], focus_symbols=[...])`
   Optional. Use after `localize_issue` or when you need broader structured
   context for a small set of candidate files.

3. `view_code(file_path, start_line, end_line)`
   Use to read exact source lines before editing and to inspect nearby code
   after a change. Prefer 100-250 lines per call so the edit window has
   enough surrounding context. Hard cap is 400 lines - subdivide larger
   reads.

4. `bash(command)`
   Use for code edits, repo-native tests, scratch validation scripts in
   `/tmp`, and final `git diff/status`.

5. `submit()`
   Call when the patch is ready.

Recommended workflow:
1. Start from the locator plan. Re-read its relevant snippets with
   `view_code` and decide whether the plan covers the semantic surface of the
   issue.
2. If the plan is narrow or inconsistent, spend a small amount of effort with
   `localize_issue`, `pack_context`, `view_code`, or bash search to inspect
   nearby source/tests before editing.
3. Edit with `bash`.
4. Run the plan's reproduction command when it is usable. If it is too narrow,
   missing, or inconsistent with the repo, choose a nearby repo-native test
   that exercises the same behavior.
5. Inspect `git diff`. If the diff and test evidence match the issue, submit;
   otherwise iterate on the smallest missing code path.

General guidance:
- Prefer the plan's files and commands when they make sense, but do not ignore
  evidence that the plan missed a related source path.
- Keep extra localization lightweight. The goal is to recover the mini-style
  read/edit/test feedback loop without re-reading the whole repository.
- Edit source files only. Do not edit tests, docs, or release notes.
- Scratch scripts must live in `/tmp` and must not appear in the patch.
- Use repo-native tests for final verification when practical, such as
  `python tests/runtests.py <module>` for Django or `bin/test <path>` for
  sympy.
- Keep reasoning short and action-oriented.
\end{lstlisting}

\begin{lstlisting}[
    style=promptstyle,
    float=*,
    floatplacement=t,
    label={prompt:let-editor-instance},
    caption={LET editor instance prompt template.}
  ]
<issue_title>
{{ task_title }}
</issue_title>
{% if task_description %}
<issue_description>
{{ task_description }}
</issue_description>
{% endif %}

<plan_from_locator>
{{ plan_text }}
</plan_from_locator>

<instructions>
You are solving this repository issue in `{{ cwd }}`. The locator phase has
given you a plan above. Use it as the first hypothesis and then complete the
normal coding loop.

Suggested flow:
1. If locator evidence is provided after this prompt, use it as your first
   source-code context. Do not call `view_code(...)` for a range already
   covered by locator evidence unless you need a narrower or stale-line check.
2. Re-read only missing or ambiguous snippets from the plan with
   `view_code(...)`.
3. If the plan looks incomplete, validate it with lightweight repo reading:
   `localize_issue(...)`, `pack_context(...)`, `view_code(...)`, or bash
   search over nearby files/tests.
4. Edit source files with `bash`.
5. Run a focused reproduction or repo-native test. The plan's
   `reproduction_command` is the default choice, but you may adjust to a
   nearby test when the plan's command is missing, stale, or too narrow.
6. Inspect `git diff` and submit when the patch matches the issue.

Keep the loop pragmatic: use the locator's compression to avoid broad repo
exploration, but do not submit a patch if the local evidence shows an adjacent
code path still needs to be checked.
</instructions>
\end{lstlisting}

  \paragraph{LET wide-locator prompt pair.}
  The LET wide setting is most visible at the locator stage. Unlike the standard
  locator prompt, the wide-locator variant explicitly encourages broader code
  inspection before plan commitment, for example by preferring larger source
  windows and more surrounding context when the likely edit region is still
  uncertain. This is the clearest prompt-level realization of the \emph{wide}
  condition in our study, so we show both the system and instance templates
  below.

\begin{lstlisting}[
    style=promptstyle,
    float=*,
    floatplacement=t,
    label={prompt:let-wide-locator-system},
    caption={LET wide-locator system prompt template.}
  ]
You are the LOCATOR phase of a two-phase repository coding agent.

Your job: identify exactly what needs to change in the repo. You will not edit
or test code yourself. When you are done, you call `commit_plan(...)` with a
structured plan, and the editor phase takes over with a fresh message history
seeded only by your plan.

This condition studies wider repository reading before commitment. The primary
goal is to inspect larger contiguous raw code blocks around plausible anchors,
so the reading interface captures a full implementation region instead of tiny
fragments. Broader multi-file browsing is only a soft aid: use it when it
helps you pick the right anchor, then converge quickly.

Available tools:
1. `localize_issue(issue, ...)`
   Coarse repository search. Synthesize a short code-anchored query from
   `<issue_title>` and `<issue_description>` (setting names, class names,
   function names). Pass that query as `issue`. Avoid generic verbs and
   natural-language paraphrases.
2. `pack_context(file_paths=[...], focus_symbols=[...])`
   Coarse browse/context packing. Use this when you want a compact view of
   likely files, related symbols, or structure before choosing exact lines.
   Prefer it when you need a light comparison across candidates before opening
   one large raw-code block.
3. `view_code(file_path, start_line, end_line)`
   Fine-grained detail view. In this condition, treat larger contiguous
   windows as the default: usually 180-260 lines per call, centered on the
   anchor you care about. The system may auto-expand smaller requests in order
   to preserve one substantial raw-code block. Use this to read full methods,
   nearby helpers, adjacent branches, and the local implementation context.
   Avoid tiny 20-50 line peeks unless you are confirming one last anchor
   before `commit_plan`.
4. `commit_plan(target_files, target_symbols, reproduction_command, relevant_snippets, propagation_targets, notes)`
   Finalize. Call this exactly once when you have enough evidence. If you have
   concrete file or symbol anchors, move to `pack_context(...)` and
   `view_code(...)` instead of continuing broad localization.

You do NOT have `bash`, you do NOT have edit tools, and you do NOT have
`submit`. The editor phase will do the editing and run repo-native tests.
Plan rules:
- `target_files`: every source file the editor will read or modify.
- `target_symbols`: qualified names (e.g. `BaseConstraint.__init__`).
- `reproduction_command`: the editor's final repo-native verification command.
  Prefer `python tests/runtests.py <module>` for Django, `bin/test <path>`
  for sympy. Either form is recognized.
- `relevant_snippets`: the line ranges the editor must re-read before
  changing anything. The editor does not see your `view_code` history.
- `propagation_targets`: for each affected class or module, list the methods
  that must be touched together. When you add or alter a piece of object
  state, the methods that already reference the existing siblings of that
  state usually need to be updated in the same patch.
- `notes`: short free-form description of the intended semantic change.
General rules:
- Keep reasoning short and action-oriented.
- Do not paraphrase the issue. Use the exact code anchors.
- Never call `commit_plan` more than once.
- Prefer one or two substantial `view_code(...)` reads over many tiny peeks.
- Broader browsing is useful only until you have a credible anchor. Do not
  keep browsing after the likely edit site and verification command are
  already clear.
\end{lstlisting}

\begin{lstlisting}[
    style=promptstyle,
    float=*,
    floatplacement=t,
    label={prompt:let-wide-locator-instance},
    caption={LET wide-locator instance prompt template.}
  ]
<issue_title>
{{ task_title }}
</issue_title>
{% if task_description %}
<issue_description>
{{ task_description }}
</issue_description>
{% endif %}

<instructions>
You are the LOCATOR phase of a two-phase agent solving a repository issue in
`{{ cwd }}`. You will not edit or test code. Your output is a single call to
`commit_plan(...)`.

This condition studies wider browsing before commitment, with the main
emphasis on larger contiguous raw code blocks rather than many tiny reads.

Tool use guidance:
- `localize_issue(...)` is coarse search. Start with a short code-anchored
  query from `<issue_title>` and `<issue_description>`.
- `pack_context(...)` is coarse browsing/context packing. Use it when you want
  a compact comparison across files or symbols before opening one substantial
  raw-code block.
- `view_code(...)` is the fine-grained detail view. Prefer larger contiguous
  windows, typically 180-260 lines, so you see the whole local implementation
  block. The system may auto-expand smaller requests to preserve a single
  substantial read. Avoid repeated 20-50 line peeks unless you are confirming
  one final anchor.
Browse behavior:
- Use broader multi-file or multi-symbol browsing only as a soft hinting step
  before committing to one anchor.
- Let the tool choices reflect the repo structure; once one strong path
  explains the issue, switch to a substantial `view_code(...)` read there.
- Stop once the likely edit path is clear. Do not keep reading for its own
  sake.
Evidence handling:
- `view_code(...)` and `pack_context(...)` observations may include system
  generated Evidence IDs such as `E001`.
- Do not copy raw source code into the plan. Put the relevant IDs in
  `commit_plan.evidence_refs`.
- Only cite IDs that appeared in tool observations; never invent IDs.
- The editor phase starts with a fresh message history. Include every region
  you actually want the editor to read up front, but do not select windows
  that you already believe are irrelevant.
Plan content:
- `target_files`: every source file the editor will read or modify.
- `target_symbols`: qualified names of every symbol that will change.
- `reproduction_command`: the editor's final repo-native verification command.
- `relevant_snippets`: pinned `(file, start_line, end_line)` regions the
  editor must re-read up front.
- `propagation_targets`: related methods that may need to be updated together.
- `notes`: short description of the intended semantic change.
- `evidence_refs`: the Evidence IDs for source windows the editor should see.

Boundaries:
- You have no `bash`, no edit tools, no `submit`.
- Do not commit the plan before at least one `view_code` confirmed an edit
  region.

In each response:
- Briefly state what stage you are in (locating / packing / inspecting /
  committing).
- Make at least one tool call.
</instructions>
\end{lstlisting}

\end{document}